\documentstyle[11pt]{article}
\def\eachbox#1#2#3#4{\lower #4\hbox{\vrule\vbox{\hrule
   \hbox to #1{\hfil\vbox to #2{\vss{\hbox{#3}}\vss
       }\hfil}%
\hrule}\vrule}\hskip-0.4pt}%
\def\onebox{\eachbox{10pt}{10pt}{}{1.5pt}}%
\def\anotherbox{\eachbox{20pt}{10pt}{$\cdots$}{1.5pt}}%

\begin{document}

\title{WORLD SPINORS REVISITED}

\author{Dj. \v{S}ija\v{c}ki \\
Institute of Physics, P.O. Box 57, 11001 Belgrade, Yugoslavia}
\date{}

\maketitle

\begin{abstract}
World spinors are objects that transform w.r.t. double covering group
$\overline{Diff}(4,R)$ of the Group of General Coordinate Transformations.
The basic mathematical results and the corresponding physical interpretation
concerning these, infinite-dimensional, spinorial representations are
reviewed.  The role of groups $Diff(4,R)$, $GA(4,R)$, $GL(4,R)$, $SL(4,R)$, 
$SO(3,1)$ and the corresponding covering groups is pointed out. New results
on the infinite dimensionality of spinorial representations, explicit
construction of the $\overline{SL}(4,R)$ representations in the basis of
finite-dimensional non-unitary $SL(2,C)$ representations, $SL(4,R)$
representation regrouping of tensorial and spinorial fields of an arbitrary
spin lagrangian field theory, as well as its $SL(5,R)$ generalization in the
case of infinite-component world spinor and tensor field theories are
presented.
\end{abstract}

\section{Introduction}

The basic wisdom of the standard approach to General Relativity is to start
with the group of "general coordinate transformations" ($GCT$), i.e. the
group of diffeomorphisms $Diff(4,R)$ of $R^4$. The theory is set upon the
principle of general covariance. The $GCT$ group has finite-dimensional
tensorial representations only, and these representations characterize
allowed world fields. A unified holonomic description of both tensors and
spinors would require the existence of respectively tensorial and (double
valued) spinorial representations of the $GCT$ group. In other words one is
interested in the corresponding single-valued representations of the double
covering $\overline{GCT}$ of the $GCT$ group, since the topology of $GCT$ is
given by the topology of its linear compact subgroup. It is well known that
the finite-dimensional representations of $\overline{GCT}$ are characterized
by the corresponding ones of the $\overline{SL}(4,R) \subset
\overline{GL}(4,R)$ group, and $\overline{SL}(4,R)$ does not have {\it
finite} spinorial representations. However, {\it there are
infinite-dimensional} $\overline{SL}(4,R)$ {\it spinorial representations}
that define the true "world" (holonomic) spinors~\cite{1}.

There are two basic ways to introduce finite spinors in a generic curved
space-time: i) One can make use of the nonlinear representations of the
$\overline{GCT}$ group, which are linear when restricted to the Poincar\'e
subgroup~\cite{2} with metric as a nonlinear realizer field. ii) One can
introduce a bundle of cotangent frames, i.e. a set of 1-forms $e^a$
(tetrads; $a=0,\ldots , 3$ the anholonomic indices) and define in this space
an action of a physically distinct local Lorentz group. Owing to this
Lorentz group one can introduce finite spinors, which behave as scalars
w.r.t. $\overline{GCT}$. The bundle of cotangent frames represents an
additional geometrical construction corresponding to the physical
constraints of a local gauge group of the Yang-Mills type, in which the
gauge group is the isotropy group of the space-time base manifold.

In order to set up a framework for a unified description of both tensors and
spinors one is now naturally led to enlarge the local Lorentz group to the
whole linear group $\overline{GL}(4,R)$, and together with translations one 
obtains the affine group $\overline{GA}(4,R)$. The affine group translates 
and deforms the tetrads of the locally Minkowskian space-time~\cite{3}, and 
provides one with either infinite-dimensional linear or finite-dimensional
nonlinear spinorial representations~\cite{4}.

The existence and structure of spinors in a generic curved space have been
the subject of more confusion than most issues in mathematical physics. The
physics literature contains two common errors: 

(i) For fifty years, it was wrongly believed that the double-covering of
$GL(n,R)$, $n \ge 3$, which we shall denote $\overline{GL}(n,R)$ does not
exist. Almost every textbook in general relativity theory, upon reaching the
subject of spinors, contains a sentence such as "... there are no
representations of $GL(4,R)$, or even "representations up to a sign", which
behave like spinors under the Lorentz subgroup". Y. Ne'eman played a pioneer
role in clarifying the issue of the double covering $\overline{SL}(n,R)
\subset \overline{GL}(n,R)$ existence~\cite{5}, and together with F.W.
Hehl~\cite{6} envisaged a gauge theory of gravity with infinite-component
spinorial matter fields. Though the correct answer has been known (and
strengthened~\cite{7}) for the last twenty years, the same type of erroneous
statement continues to appear in more recent texts. The complete list of the
(infinite-dimensional) $\overline{SL}(3,R)$ and $\overline{SL}(4,R)$ unitary
irreducible representations is known~\cite{8,9}, a formulation of
(super-symmetric) spinning extended objects in a generic curved space is
developed~\cite{10}, as well as Gauge Affine and Metric Affine Gauge
Theories of Gravity with tensor and spinor $\overline{GL}(4,R)$ matter
fields have been developed considerably~\cite{11,12}.

(ii) An additional reason for the overall confusion concerns the unitarity
of the relevant spinor representations. In dealing with non-compact groups,
it is customary to select infinite-dimensional unitary representations to
describe the particle-states. However, in the standard (point-object) field
theory of tensors or spinors the finite,  non-unitary representations of
$GL(4,R)$ and $SL(2,C)$ are used respectively. The correct answer for
spinorial $\overline{GL}(4,R)$ fields consists in using the infinite unitary
representations in a physical base in which they become non-unitary for the
$SL(2,C)$ subgroup~\cite{13}. In this way one describes the experimental
facts that elementary particles (say proton) when boosted do not turn into
another particles (hadronic states) of the same infinite-component spinorial
field. Field equations have been constructed for such infinite-component
fields, "manifields", within Riemannian gravitational theory~\cite{14,1}.
$\overline{SL}(4,R)$ manifields have also been used in classifying the
hadron spectrum~\cite{15}.

\section{World Spinors Existence}

Let $g_{0}=k_{0}+a_{0}+n_{0}$ be an Iwasawa decomposition of a semi-simple
Lie algebra $g_{0}$ over $R$. Let $G$ be any connected Lie group with Lie
algebra $g_{0}$, and let $K$ (compact), $A$ (Abelian) and $N$ (nilpotent) be
the analytic subgroups of $G$ with Lie algebras $k_{0}$,$a_{0}$ and $n_{0}$
respectively. The mapping $(k,a,n) \rightarrow kan\quad (k\in K,a\in A,n\in
N)$ is an analytic diffeomorphism of the product manifold $K\times A\times
N$ onto $G$. The groups $A$ and $N$ are simply connected. Only $K$ is not
guaranteed to be simply-connected. There exists a universal covering group
$\overline{K}$ of $K$, and thus also a universal covering $\overline{G}
\simeq \overline{K} \times A \times N$ of $G$. For the group of
diffeomorphisms one has the following decomposition
$$
Diff(n,R) = GL(n,R) \times H \times R^{n} 
$$
where the subgroup $H$ is contractible to a point. As $O(n)$ is the compact
subgroup of $GL(n,R)$, one finds that $O(n)$ is a deformation retract of
$Diff(n,R)$. Thus, there exists a universal covering of the Diffeomorphism
group
$$
\overline{Diff}(n,R) \simeq \overline{GL}(n,R) \times H \times R^{n}. 
$$
Summing up, we note that for $n \ge 3$ both $SL(n,R)$ and on the other hand
$GL(n,R)$ and $Diff(n,R)$ will all have double coverings, defined by
$\overline{SO}(n) \simeq Spin(n)$ and $\overline{O}(n) \simeq Pin(n)$ 
respectively, the double-coverings of the $SO(n)$ and $O(n)$ maximal compact 
subgroups.

We have proven previously~\cite{7} that $\overline{SL}(4,R)$ cannot be
embedded into either $SL(4,C)$ or any other classical semi-simple Lie group.
Here we demonstrate on the simplest $\overline{SL}(3,R)$ example how infinite
matrices appear. Let $J_i$ ($i=1,2,3$; angular momentum) and $T_k$ ($k=1,
\ldots ,5$; shear) be the $\overline{SL}(3,R)$ generators. For the simplest
(multiplicity free) representations, one obtains in the spherical basis the
following reduced matrix elements of the non-compact (shear) 
generators~\cite{8} 
\begin{eqnarray*}
<j-2 ||T || j> = 
-i(-)^{2j} (\sigma_{1} + i\sigma_{2} + 2j - 1) 
\sqrt{\frac{j(j-1)}{2j-1}}, \\
<j || T || j> = +i(-)^{2j} (\sigma_{1} + i\sigma_{2}) 
\sqrt{\frac{2j(j+1)(2j+1)}{3(2j+3)(2j-1)}}, \\
<j+2 || T || j> = -i(\sigma_{1} + i\sigma_{2} - 2j - 3) 
\sqrt{\frac{(j+1)(j+2)}{2j+3}}. \\
\end{eqnarray*}
where $\sigma_{1}, \sigma_{2} \in R$. One can have angular momentum
$j = 1/2$ provided $\sigma_{1} = \sigma_{2} = 0$
($<j_{min}-2||T||j_{min}>=0$, $j_{min}=1/2$), however in this case one obtains all $j = 5/2, 
9/2, \ldots$ as well, and an infinite non-compact matrix for the shear
generators.

\section{General affine particles and world spinor fields}

The finite-dimensional world tensor field components are characterized by
the non-unitary representations of the homogeneous group $GL(4,R)$ $\subset$
\hfill\break 
$Diff(4,R)$. In the flat-space limit they split up into non-unitary $SL(2,C)$
irreducible pieces. The particle states are defined in the tangent
flat-space only. They are characterized by the unitary irreducible
representations of the (inhomogeneous) Poincar\'e group $P(4) = T_4\wedge
SL(2,C)$, and they are enumerated by the "little" group unitary
representations (e.g. $T_3\otimes SU(2)$ for $m\ne 0$). In the generalization
to world spinors, the $SL(2,C)$ group is enlarged to the $\overline{SL}(4,R)
\subset \overline{GL}(4,R)$ group, while $GA(4,R) = T_4\wedge
\overline{GL}(4,R)$ is to replace the Poincar\'e group. Affine "particles"
are characterized by the unitary irreducible representations of the
$\overline{GA}(4,R)$ group, whose unitarity is provided by the unitarity of
the relevant "little" group (e.g. $T_3\otimes \overline{SL}(3,R) \supset
T_3\otimes SU(2)$). A mutual particle--field matching is achieved by
requiring the subgroup of the homogeneous group, that is isomorphic to the
homogeneous part of the "little" group (say, $SU(2)$ of $SL(2,C)$), to be
represented unitarily. Furthermore, one has to project away all
representations of this group except a single one that is realized for the
particle states (say $D^{(j)}$ of  $SU(2)\subset T_3\otimes SU(2)$). 

A physically correct picture, in the affine case, is obtained by making use
of the $\overline{SA}(4,R) \subset \overline{GA}(4,R)$ group unitary
irreducible representations for "affine" particles, with particular states
characterized by the $T_3\otimes \overline{SL}(3,R)$ "little" group
representations. The corresponding affine fields are described by the
non-unitary infinite-dimensional $\overline{SL}(4,R) \subset
\overline{GL}(4,R)$ representations, that are unitary when restricted to the
homogeneous "little" subgroup $\overline{SL}(3,R)$. Therefore, the first
step towards world spinor fields is a construction of infinite-dimensional
non-unitary $\overline{SL}(4,R)$ representations, that are unitary when
restricted to $\overline{SL}(3,R)$. These fields reduce to an infinite sum
of (non-unitary) finite-dimensional $SL(2,C)$ fields. 

The world spinor fields transform w.r.t. $\overline{Diff}(4,R)$ as follows
$$
(D(a,\bar f)\Phi_A) (x) = (D_{\{ \overline{Diff_0}\} })^B_A (\bar f)\Phi_B 
(f^{-1}(x-a)),\quad (a,\bar f) \in T_4 \wedge \overline{Diff_0},
$$
where $\overline{Diff_0}$ is the homogeneous part of $\overline{Diff}$, and 
$D_{\{ \overline{Diff_0} \} } = {\sum}^\oplus D_{\{ \overline{SL} \} }$. The
affine "particle" states transform according to the following representation
$$
D\big( (a, \bar s)\big) \rightarrow e^{i(sp)\cdot a} D_{\{ \overline{SL}\} }
(L^{-1}(sp)\bar s L(p)) ,\quad (a,\bar s) \in T_4\wedge
\overline{SL}(4,R),
$$
and $L \in \overline{SL}(4,R)/\overline{SL}(3,R)$ The unitarity properties
of various representations in these expressions is as described above.

\section{Spinorial $SL(2,C) \subset \overline{SL}(4,R)$ representations}

In order to analyze the representations, as well as to make use of them in a
gauge theory, it is convenient to have the matrix elements of the group
generators. Also, in that case the task of determining the scalar products
of the irreducible representations is considerably simplified. Let
$M_{\mu\nu}$ and $T_{\mu\nu}$ be the $\overline{SL}(4,R)$ generators, with
$M_{\mu\nu}$ generating the Lorentz subgroup $SL(2,C) \simeq
\overline{SO}(3,1)$. In the $3+1$ notation one has $M_{\mu\nu}$
$\rightarrow$ $J_i = \epsilon_{ijk}M_{jk}$ (angular momentum), $K_i =
M_{0i}$ (boost), and $T_{\mu\nu}$ $\rightarrow$ $T_{ij}$ ($3$-shear), $N_i =
T_{0i} = T_{i0}$, and $T_{00}$. At this point it is convenient (as in the
Lorentz covariant field theory) to introduce $J^{(1)}_i$, and $J^{(2)}_i$
that generate an $SU(2)\otimes SU(2)$ group with the corresponding
representation labels $(j_1 , j_2)$, $j_1,\ j_2=0,\ 1/2,\ 1\ ...$. 

The angular momentum and boost generators are given by $J_i = J^{(1)}_i + 
J^{(2)}_i$, and $K_i = i(J^{(2)}_i - J^{(1)}_i)$. The remaining 
$\overline{SL}(4,R)$ generators transform as a $(1,1)$ $SU(2)\otimes SU(2)$ 
irreducible tensor operator $Z_{pq},\ p,q = 0, \pm 1$. The most general 
$\overline{SL}(4,R)$ representations are obtained in the $\left| 
\begin{array}{cc}j_1 & j_2 \\ k_1\ m_1 & k_2\ m_2 \end{array} \right>$ basis
of the $SU(2)\otimes SU(2)$ representations. In the reduction to the
$SL(2,C)$ subgroup they contain an infinite direct sum of corresponding
irreducible representations $D^{(j_1 , j_2 )}_{SL(2,C)}$.

The matrix elements of the Lorentz group generators are well known, and we
list only the matrix elements of the $Z_{pq}$ generators obtained by the
appropriate generator redefinition as compared to $\overline{SL}(4,R) /
\overline{SO}(4)$ representations~\cite{9}.

$
\left< \begin{array}{cc} j_1^\prime & j_2^\prime \\ k_1^\prime\ m_1^\prime &
k_2^\prime\ m_2^\prime \end{array} \right| 
Z_{pq} 
\left| \begin{array}{cc} j_1 & j_2 \\ k_1\ m_1 & k_2\ m_2 \end{array}\right> 
 = (-)^{j_1^\prime -m_1^\prime}(-)^{j_2^\prime-m_2^\prime} 
$\hfill
$$
\times 
\left( \begin{array}{ccc} j_1^\prime & 1 &j_1 \\ -m_1^\prime & p & m_1
\end{array} \right)
\left(\begin{array}{ccc} j_2^\prime & 1 & j_2 \\ -m_2^\prime & q & m_2 
\end{array}\right)
\left<\left.  \begin{array}{cc} j_1^\prime & j_2^\prime \\  k_1^\prime & 
k_2^\prime \end{array}
\right|\right| Z \left|\left| \begin{array}{cc} j_1 & j_2 \\ k_1 & k_2 
\end{array}\right>\right. ,
$$
where,

$
\left<\left. \begin{array}{cc} j_1^\prime & j_2^\prime \\  k_1^\prime & 
k_2^\prime \end{array}
\right|\right| Z \left|\left| \begin{array}{cc} j_1 & j_2 \\ k_1 & k_2 
\end{array}\right>\right. 
$\hfill

$= (-)^{j_1^\prime-k_1^\prime}(-)^{j_2^\prime -k_2^\prime}\frac{i}{2}
\sqrt{(2j_1^\prime +1)(2j_2^\prime +1)(2j_1+1)(2j_2+1)} $

$\times \Biggl\{ [e+4-j_1^\prime (j_1^\prime +1)+j_1(j_1+1)-j_2^\prime
(j_2^\prime +1)+j_2(j_2+1)]$

$\times
\left( \begin{array}{ccc} j_1^\prime & 1 &j_1 \\ -k_1^\prime & 0 & k_1 
\end{array}\right)
\left( \begin{array}{ccc} j_2^\prime & 1 &j_2 \\ -k_2^\prime & 0 & k_2 
\end{array} \right)$

$-(c+k_1-k_2)
\left( \begin{array}{ccc} j_1^\prime & 1 &j_1 \\ -k_1^\prime & 1 & k_1 
\end{array} \right)
\left( \begin{array}{ccc} j_2^\prime & 1 &j_2 \\ -k_2^\prime & -1 & k_2 
\end{array} \right)$

$-(c-k_1+k_2)
\left( \begin{array}{ccc} j_1^\prime & 1 &j_1 \\ -k_1^\prime & -1 & k_1
\end{array} \right)
\left(\begin{array}{ccc} j_2^\prime & 1 &j_2 \\ -k_2^\prime & 1 &k_2 
\end{array} \right)$

$+(d+k_1+k_2)
\left( \begin{array}{ccc} j_1^\prime & 1 &j_1 \\ -k_1^\prime & 1 & k_1
\end{array} \right)
\left( \begin{array}{ccc} j_2^\prime & 1 &j_2 \\  -k_2^\prime & 1 &k_2 
\end{array} \right)$

$+(d-k_1-k_2)
\left( \begin{array}{ccc} j_1^\prime & 1 &j_1 \\ -k_1^\prime & -1 & k_1
\end{array} \right)
\left( \begin{array}{ccc} j_2^\prime & 1& j_2 \\ -k_2^\prime & -1 &k_2 
\end{array} \right)\Biggr\},$

\noindent The representation labels $c$, $d$, $e$ are arbitrary complex
numbers.

A class of infinite-dimensional spinorial/tensorial representations of the
$GL(4,R)$ group in the basis of its Lorentz subgroup were recently
constructed~\cite{16} by extending the $SL(2,C)$ Naimark representations.
These representations are made up of infinite-dimensional $SL(2,C)$
representations and fail to meet the necessary physical requirements.

\section{Minimal field configurations for arbitrary spin lagrangian theory}

The task of constructing a lagrangian formulation for relativistic fields of
unique mass and arbitrary spin turns out to be rather non-trivial. There is
no unique Lorentz covariant field to be associated to a given $[m,J]$
particle Poincar\'e representation. Moreover, as a rule, lagrangian
formulation requires quite a number of additional auxiliary fields. 

A minimal Lorentz covariant Fierz-Pauli lagrangian formulation of an massive 
arbitrary-spin $s$ boson field is achieved in terms of traceless, symmetric
tensor fields~\cite{17, 18}. Let $\phi^{(s)}$ be a Lorentz covariant field
that transforms w.r.t. the $D^{(\frac{s}{2},\frac{s}{2})}$ $SL(2,C)$ 
irreducible representation (a symmetric traceless tensor of rank $s$) 
satisfying the Klein-Gordon equation with mass $m$, i.e. $(\onebox + m^2) 
\phi^{(s)}_{\nu_1\nu_2 \cdots\mu_s}(x) = 0$. Representation 
$D^{(\frac{s}{2},\frac{s}{2})}$ is reducible under the $SO(3)$ subgroup of
spatial rotations, $D^{(\frac{s}{2},\frac{s}{2})}$ $=$
$\sum_{l=0}^{s} D^{(l)}$, and thus one imposes the "Lorentz condition"
$\partial^{\mu_1} \phi^{(s)}_{\mu_1\mu_2\cdots\nu_s} = 0$ in order to
eliminate all lower spin values. In order to have enough field components to
vary, it is necessary to introduce certain auxiliary fields. The simplest
viable choice is to introduce, besides the starting field $\phi^{(s)}$, the
following set of auxiliary fields: $\phi^{(s-2)},\ \phi^{(s-3)},\ \cdots\ ,  
\phi^{(0)}$. The total field is $\Phi^{(s)}$ $=$ $\{ \phi^{(s)},\  
\phi^{(s-2)},\ \phi^{(s-3)},\ \cdots\ \phi^{(0)}\}$, and consists of $(s+1)^2 
+ \frac{1}{6}s(s+1)(2s-1)$ field components. 

The fields of a generic curved-space theory formulation transform w.r.t.
linear $SL(4,R) \subset GL(4,R)$ representations, that provide a space for
non-linear realization of the complete $Diff(4,R)$ transformation group.
Therefore, the first step in formulating a generic curved-space lagrangian
field theory for arbitrary spin is to embed the space of all fields of
the above minimal Lorentz formulation into an appropriate $SL(4,R)$
representation space. An analysis of $SL(2,C)$ and $SL(4,R)$ representations
shows that the basic field $\phi^{(s)}$ as well as all accompanying auxiliary
fields can be reorganized to fit into two $SL(4,R)$ irreducible 
representations when $s \ge 3$, while for $s = 0, 1, 2$ a single $SL(4,R)$
representation suffice. In the Young tableau notation of $SL(4,R)$
irreducible representations, we find 
$$
\Phi^{(s)} \quad\sim\quad \underbrace{\onebox\onebox\anotherbox\onebox}_{s} 
\quad\bigoplus\quad \underbrace{\onebox\onebox\anotherbox\onebox}_{s-3} 
$$
when $s \ge 3$, while $\Phi^{(0)} \sim \bullet$ (scalar representation; also
the second representation when $s=3$), $\Phi^{(1)} \sim \onebox$\ , and 
$\Phi^{(2)} \sim \onebox\onebox$ . 

A minimal Lorentz covariant Fierz-Pauli lagrangian formulation of an massive
arbitrary-spin $j = \frac{1}{2} + s$ fermion field is achieved in terms of
Rarita-Schwinger spinor-tensor fields~\cite{17, 18}. Let $\psi^{(s)}$ be a
symmetric traceless spinor-tensor field that transforms w.r.t. 
$D^{(\frac{1}{2}(s+1),\frac{1}{2}s)}$ $\oplus$ 
$D^{(\frac{1}{2}s,\frac{1}{2}(s+1))}$
representation of the $SL(2,C)$ group and satisfies $(i\gamma\cdot\partial -
m)\psi^{(s)} (x) =0$, and the spinor trace condition $\gamma^{\mu_1} 
\psi^{(s)}_{\mu_1\mu_2\cdots\mu_s}(x) = 0$. The lagrangian formulation is
achieved for a field $\Psi^{(s)}$ $=$ $\{ \psi^{(s)},\ \psi^{(s-1)},\
2\times\psi^{(s-2)},\ 2\times\psi^{(s-3)}, \cdots 2\times\psi^{(0)} \}$,
transforming w.r.t. 
$D^{(\frac{1}{2}(s+1),\frac{1}{2}s)}$ $\oplus$ 
$D^{(\frac{1}{2}s,\frac{1}{2}(s+1))}$ 
$\oplus$ 
$D^{(\frac{1}{2}s,\frac{1}{2}(s-1))}$ $\oplus$ 
$D^{(\frac{1}{2}(s-1),\frac{1}{2}s)}$ 
$\oplus$ $2\ \sum_{l=0}^{s-2}$ 
$[D^{(\frac{1}{2}(l+1),\frac{1}{2}l)}$ $\oplus$ 
$D^{(\frac{1}{2}l,\frac{1}{2}(l+1))}]$. 
representation of the Lorentz group, and consists of the starting field
$\psi^{(s)}$ and the necessary auxiliary fields.

In this case, we find that the tensor part of the spinor-tensor field can,
again, be described by two $SL(4,R)$ irreducible representations. In the
Young tableau notation we write 
$$
\Psi^{(s)}\quad\sim\quad [\ D^{(\frac{1}{2},0)}\bigoplus 
D^{(0,\frac{1}{2})}\ ] \bigotimes 
[\ \underbrace{\onebox\onebox\anotherbox\onebox}_{s} \bigoplus 
\underbrace{\onebox\onebox\anotherbox\onebox}_{s-2}\ ]
$$
when $s \ge 2$, while $\Psi^{(0)} \sim [D^{(\frac{1}{2},0)}\oplus 
D^{(0,\frac{1}{2})}]\otimes\bullet$, and $\Psi ^{(1)}$ $\sim$ 
$[D^{(\frac{1}{2},0)}\oplus D^{(0,\frac{1}{2})}]$ 
\hfill\break $\otimes\onebox$ . 
Here, the spinor and tensor parts transforms w.r.t. $SL(2,C)$ and $SL(4,R)$
representations respectively. 

\section{World spinor field choice and $\overline{SL}(5,R)$}

Let us consider world tensor and spinor fields that transform according to
infinite-dimensional $\overline{SL}(4,R)$ representations that consists of
finite-dimen\-si\-o\-nal, non-unitary $SL(2,C)$ subgroup representations.
Owing to the fact that each infinite-dimensional $\overline{SL}(4,R)$
representation contains an infinite set of Lorentz representations, i.e. an
infinite set of tensors or spinors, one has a structure that should contain
$\sum_{s=0}^{\infty} \Phi^{(s)}$ or $\sum_{s=0}^{\infty} \Psi^{(s)}$ fields
at least. 

The simplest tensorial case is based on the multiplicity-free (ladder)
$SL(4,R)$ representations~\cite{13} $D^{ladd}_{SL(4,R)}(0,0)$ and 
$D^{ladd}_{SL(4,R)}(\frac{1}{2},\frac{1}{2})$ that contain
each $D^{(\frac{s}{2},\frac{s}{2})}$, $s = 0,1,\ldots$, $SL(2,C)$
representation once. However, each Lorentz covariant field $\phi^{(s)}$ is
accompanied by the appropriate auxiliary fields: $\phi^{(s)} \rightarrow
\Phi^{(s)}$, resulting in an infinite repetition of the starting $SL(4,R)$
representations. One obtains a structure resembling that of the leading and
daughter Regge trajectories of hadrons. We find that {\it all these field
components} (physical and auxiliary) {\it can be obtained from a single
infinite-dimensional $SL(5,R)$ multiplicity-free representation.}
$$
\tilde{\Phi} \quad\sim\quad D^{(ladd)}_{SL(5,R)} \quad\supset\quad 
{\sum}^{\oplus} [D^{(ladd)}_{SL(4,R)}(0,0) \bigoplus
D^{(ladd)}_{SL(4,R)}(\frac{1}{2},\frac{1}{2})]
$$

We find, in the spinor field case, an analogous result. For a spinor-tensor
field $\tilde{\Psi}$ that transforms as a Dirac field w.r.t. the Lorentz
group, and as a tensor w.r.t. $SL(4,R)$, we have
$\tilde{\Psi} \sim [D^{(\frac{1}{2},0)}\oplus D^{(0,\frac{1}{2})}]\otimes 
D^{(ladd)}_{SL(5,R)}$ $\supset$ 
$\sum^{\oplus} [D^{(\frac{1}{2},0)}\oplus D^{(0,\frac{1}{2})}]\otimes 
[D^{(ladd)}_{SL(4,R)}(0,0) \oplus
D^{(ladd)}_{SL(4,R)}(\frac{1}{2},\frac{1}{2})]$.  Finally, in order to
obtain a genuine world spinor field transforming according to the
$\overline{SL}(4,R)$ $\subset$ $\overline{Diff}(4,R)$ group spinorial
representation, we make the appropriate changes and find
$$
\tilde{\Psi} \quad\sim\quad 
D^{(ladd)}_{\overline{SL}(5,R)} \quad\supset\quad 
{\sum}^{\oplus} 
[D^{(ladd)}_{\overline{SL}(4,R)}(\frac{1}{2},0) \bigoplus 
D^{(ladd)}_{\overline{SL}(4,R)}(0,\frac{1}{2})]\ .
$$

\end{document}